\pdfoutput=1
\documentclass[runningheads]{llncs}
\usepackage[T1]{fontenc}
\usepackage{graphicx}
\usepackage{amssymb}
\usepackage{amsmath}
\usepackage{hyperref}
\usepackage{siunitx}
\newcommand{\presynaddr}{a} 
\newcommand{\postsynaddr}{b} 
\newcommand{\numevent}{N_{ev}} 
\newcommand{\presynaddrspace}{\mathcal{A}} 
\newcommand{\postsynaddrspace}{\mathcal{B}} 
\newcommand{\arank}{r} 
\newcommand{\synapse}{\mathcal{S}} 
\newcommand{\synapticweight}{w} 
\newcommand{\synapticdelay}{\delta} 
\newcommand{\ranksyn}{s} 
\newcommand{\Nsyn}{N_{s}} 
 
\newcommand{\timev}{t} 
\newcommand{\event}{\epsilon} 
\newcommand{\ms}{\si{\milli\second}}%
\newcommand{\kernel}{K} 

\newcommand{\fig}[1]{Fig.~\ref{fig:#1}}

\begin{document}

\title{
Accurate Detection of Spiking Motifs 
by Learning Heterogeneous Delays of a Spiking Neural Network
\thanks{Supported by A*MIDEX grant AMX-21-RID-025 ``\href{https://laurentperrinet.github.io/grant/polychronies/}{Polychronies}''.}
}

\author{
Laurent U Perrinet\orcidID{0000-0002-9536-010X}
}
\authorrunning{LU Perrinet}
%
\institute{INT UMR7289, Aix Marseille Univ, CNRS; 27 Bd Moulin, 13005 Marseille, France
\url{https://laurentperrinet.github.io/publication/perrinet-23-icann/} \\
\email{laurent.perrinet@univ-amu.fr}
}
\maketitle              
\begin{abstract} 
  Recently, interest has grown in exploring the hypothesis that neural activity conveys information through precise spiking motifs. To investigate this phenomenon, various algorithms have been proposed to detect such motifs in Single Unit Activity (SUA) recorded from populations of neurons. In this study, we present a novel detection model based on the inversion of a generative model of raster plot synthesis. Using this generative model, we derive an optimal detection procedure that takes the form of logistic regression combined with temporal convolution. A key advantage of this model is its differentiability, which allows us to formulate a supervised learning approach using a gradient descent on the binary cross-entropy loss. To assess the model's ability to detect spiking motifs in synthetic data, we first perform numerical evaluations. This analysis highlights the advantages of using spiking motifs over traditional firing rate based population codes. We then successfully demonstrate that our learning method can recover synthetically generated spiking motifs, indicating its potential for further applications. In the future, we aim to extend this method to real neurobiological data, where the ground truth is unknown, to explore and detect spiking motifs in a more natural and biologically relevant context.
\end{abstract}

\section{Introduction}

\subsection{The age of large-scale neurobiological event-based data}
%
Over the past decade, remarkable technological progress across multiple disciplines has expanded the potential for experimental neuroscience research. These cutting-edge methods, such as \textit{in vivo} two-photon imaging, large population recording arrays, optogenetic circuit control tools, transgenic manipulations, and large volume circuit reconstructions, allow researchers to explore neural networks' function, structure, and dynamics with unparalleled precision.

The complexity revealed by these advanced technologies underscores the significance of neurobiological knowledge in bridging the gap between abstract brain function principles and their biological implementation in neural circuits. Consequently, there is a growing need to scale up analysis methods to handle the vast amounts of data generated by these powerful techniques. By meeting this demand, researchers can gain deeper insights into brain function, further our understanding of neural circuits, and make groundbreaking discoveries in neuroscience.

One approach aimed at addressing this challenge is the Rastermap algorithm~\cite{pachitariu_robustness_2018}. This algorithm rearranges neurons in the raster map based on the similarity of their activity and utilizes a deconvolution strategy with a linear model. However, it's worth noting that the Rastermap algorithm's primary testing has been on calcium imaging data, which may introduce some imprecision in the timing of spiking activity observed in Single Unit Activity (SUA) recordings. 
Another significant contribution is from the work of Williams {\it et al.}~\cite{williams_point_2020}. They propose a point process model that overcomes limitations present in existing models, such as the need for discretized spike times or lack of uncertainty estimates for model predictions and estimated parameters. By incorporating learnable time-warping parameters to model sequences of varying durations, the model effectively captures experimentally observed patterns in neural circuits. 
%
\subsection{Decoding neural activity using spike distances}
%
Neuroscience research heavily relies on defining appropriate metrics to compute the distance between spike trains, and one well-known measure for this purpose is the Victor-Purpura distance~\cite{victor_nature_1996}. This metric effectively addresses inconsistencies observed with firing rate-based estimation of spike trains. Another study refines the Victor-Purpura distance by introducing a time constant as a parameter, allowing for interpolation between a coincidence detector and a rate difference counter~\cite{van_rossum_novel_2001}. Additionally, researchers have extended these distance measures to non-Euclidean metrics and morphological manipulations, enabling the computation of spike train dissimilarity. 

Regarding spike timings, various methods have been developed to estimate the latency of neural responses. Bayesian binning~\cite{levakova_review_2015} is one such method. Unitary event analysis, based on a statistical model of chance detection, has been widely used to detect significant synchronous patterns above chance in neuron pair recordings~\cite{grun_unitary_2002-1}. 
Recent extensions of these methods, such as the 3D-SPADE approach~\cite{stella_3d-spade_2019}, enable the identification of reoccurring patterns in parallel spike train data and assess their statistical significance. Incorporating possible temporal dithering in spike timings has been shown to improve performance, particularly in the presence of patterns with varying durations, such as surrogates used to evaluate precisely timed higher-order spike correlations. 

However, some of these methods may suffer from computational complexity, block-based implementations, and narrow specialization for specific tasks. To address these challenges, novel methods like {\sc SpikeShip}~\cite{sotomayor-gomez_spikeship_2021} are being developed. The complexity and diversity of these spike train distance and timing comparison methods demonstrate the growing interest in integrating such measures to understand the neural code. A critical step in testing their potential usefulness is scaling these methods to handle larger amounts of data, enabling broader applications and deeper insights into neural activity patterns and their significance.
\subsection{A novel hypothesis: spiking motifs}
%
\begin{figure}
  \centering
  \includegraphics[width=0.98\linewidth]{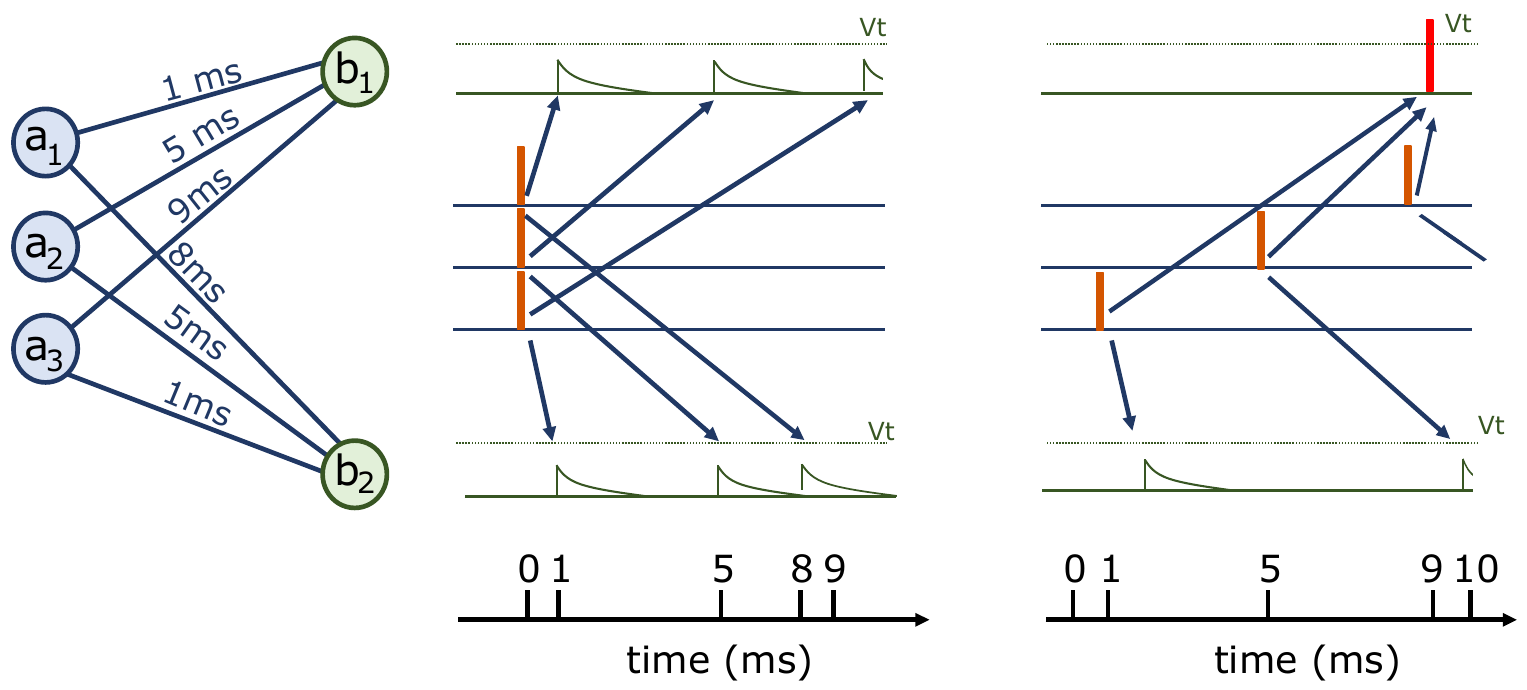}
    \caption{
      \textbf{Core Mechanism of Spiking Motif Detection:}
      In this illustrative example, we consider a scenario involving three presynaptic neurons denoted as $a_1$, $a_2$, and $a_3$, which are fully connected to two postsynaptic neurons $b_1$ and $b_2$. The synaptic delays for the connections to $b_1$ are $1$, $5$, and $9~\ms$, while for $b_2$ they are $8$, $5$, and $1~\ms$, respectively.
      In the middle panel, when the three presynaptic neurons emit synchronous pulses, the postsynaptic potentials generated in $b_1$ and $b_2$ reach them asynchronously due to the heterogeneous delays. Consequently, the postsynaptic potentials may not be sufficient to reach the membrane threshold (dashed line) in either of the postsynaptic neurons, and no output spike is generated.
      In the right panel, the pulses emitted by the presynaptic neurons are arranged in such a way that, taking into account the delays, they reach the postsynaptic neuron $b_1$ at the same time (at $t=10~\ms$ in this example). As a result, the postsynaptic potentials $V_t$ evoked by the three presynaptic neurons sum up, causing the voltage threshold to be crossed. This leads to the emission of an output spike, signaling the detection of a spiking motif in the presynaptic population (highlighted in red color).
      This core mechanism illustrates how the interplay between heterogeneous delays in the network allows for precise spike timing, enabling the detection of spiking motifs in neural populations. 
    }
  \label{fig:izhikevich}
\end{figure}
%
In recent studies, the importance of spike timing has been emphasized, especially in the barn owl auditory system, where precise spike timing in response to the sound of a mouse allows the brain to determine the prey's position~\cite{goodman_spike-timing-based_2010}. This discovery aligns with a growing body of literature suggesting that the brain's dynamics often exhibit stereotyped sequences known as \emph{spiking motifs}~\cite{grimaldi_precise_2023}. The concept of spiking motifs is a generalization of the patterns observed in the \textit{polychronization} model developed by Izhikevich~\cite{izhikevich_polychronization_2006}. This theoretical model comprises a random recurrent network of spiking neurons with biologically realistic synaptic delays and evolving weights governed by Spike-Time Dependent Plasticity (STDP) learning rule.

The interplay between the synaptic delays and STDP leads to the spontaneous organization of neurons into groups called "polychronous groups." Despite neurons in one of these groups firing at different times, the heterogeneous delays enable their spikes to converge synchronously on the postsynaptic neuron. This convergence results in the summation of excitatory postsynaptic potentials, leading to the firing of the postsynaptic neuron (see Figure~\ref{fig:izhikevich}). The polychronization model allows spiking neurons to self-organize into groups and generate reproducible time-locked spiking motifs. The STDP rule increases synaptic weights selectively for neurons involved in these polychronous groups, thereby consolidating the formation of such groups.

While the polychronization model provides valuable insights into understanding spiking neural networks and their potential role in generating spatio-temporal spiking motifs, it has a limitation. The model's heterogeneous delays are fixed and cannot evolve over time, which may limit its applicability in certain scenarios. However, the underlying mechanism offers valuable implications for studying neural activity motifs and their significance in the brain. To effectively detect spiking motifs, we propose a novel metric inspired by this model.
\subsection{The Heterogeneous Delays Spiking Neural Network (HD-SNN)}

In this work, we propose to accurately detect spatio-temporal spiking motifs using a feed-forward, single layer heterogeneous delays spiking neural network (HD-SNN). The paper is organized as follows. We develop a theoretically defined HD-SNN for which we can attune both the weights and delays. We first detail the methodology by defining the basic mechanism of spiking neurons that utilize heterogeneous delays. 
This will allow us to formalize the spiking neuron used to learn the model's parameters in a supervised manner and test its effectiveness. In the results section, we will first evaluate the efficiency of the learning scheme. We will also study the robustness of the spiking motif detection mechanism and in particular its resilience to changing the dimensions of the presynaptic or postsynaptic populations, or the depth in the number of different possible delays. Then, we will explore how the spiking motifs may be learned using supervised learning, and evaluate how the efficiency of the algorithm may depend on the parameters of the HD-SNN architecture. This will allow us to show how such a model can provide an efficient solution which may in the future be applied to neurobiological data.  Finally, we will conclude by highlighting the main contributions of this paper, while defining some limitations which will open perspectives for future detection methods. 
\section{Methods}
\label{sec:methods}
Let us formally define the HD-SNN model. First, we will define raster plots similar to those obtained from Single Unit Activity (SUA) recordings using an event-based and then binarized setting. We will then derive a generative model for raster plots using a HD-SNN, and derive a model for efficient detection of event-based motifs using a similar HD-SNN with ``inverted'' delays.
\subsection{Raster plots: from event-based to binarized}
In neurobiological recordings, 
any generic raster plot consists of a stream of \emph{spikes}. This can be formalized as a list of neural addresses and timestamps tuples $\event = \{(\presynaddr_\arank, \timev_\arank)\}_{\arank \in [1,\numevent]}$ where $\numevent \in \mathbb{N}$ is the total number of events in the data stream and the rank $\arank$ is the index of each event in the list of events. Each event has a time of occurrence $\timev_\arank$ (these are typically ordered) and an associated address $\presynaddr_\arank$ in the space $\presynaddrspace$ of the neural population. In a neurobiological recording like that of SUAs, this can be the identified set of neurons.

Events are generated by neurons which are defined on the one hand by the equations governing the evolution of its membrane potential dynamics on their soma and on the other hand by the integration of the synaptic potential propagating on their dendritic tree. A classical characterization consists in detailing the synaptic weights of each synaptic contact, the so-called weight matrix. As we saw above, neurons can receive inputs from multiple presynaptic neurons with heterogeneous delays. These delays represent the time it takes for a presynaptic spike to reach the soma of the postsynaptic neuron. 
In such neurons, 
input presynaptic spikes $\event$ will be multiplexed in time by the dendrites defined by this synaptic set (see Figure~\ref{fig:izhikevich}). 

Let's formalize such a layer of spiking neurons in the HD-SNN model. Each postsynaptic neuron $\postsynaddr \in \postsynaddrspace$  connects to presynaptic neurons from a set of addresses in  $\presynaddrspace$. In biology, a single cortical neuron has generally several thousands of synapses. Each may be defined by its synaptic weight and also its delay. 
Note that two neurons may contact with multiple synapses, and thus different delays. Scanning all neurons $\postsynaddr$, we thus define the set of $\Nsyn \in \mathbb{N}$ synapses  as  $\synapse = \{(\presynaddr_\ranksyn, \postsynaddr_\ranksyn, \synapticweight_\ranksyn, \synapticdelay_\ranksyn)\}_{\ranksyn \in [1,\Nsyn]}$, where each synapse is associated to a presynaptic address $\presynaddr_\ranksyn$, a postsynaptic address $\postsynaddr_\ranksyn$,  a weight $\synapticweight_\ranksyn$, and a delay $\synapticdelay_\ranksyn$. 

This defines the full connectivity of the HD-SNN model. The receptive field of a postsynaptic neuron refers to the set of synapses that connect to it. Similarly, the emitting field of a presynaptic neuron refers to the set of synapses it connects to. These fields determine the synaptic inputs and outputs of individual neurons. More formally, the receptive field of a postsynaptic neuron is defined $\synapse^\postsynaddr =  \{(\presynaddr_\ranksyn, \postsynaddr_\ranksyn, \synapticweight_\ranksyn, \synapticdelay_\ranksyn) \| \postsynaddr_\ranksyn=\postsynaddr\}_{\ranksyn \in [1,\Nsyn]} $, and the emitting field of a presynaptic neuron as $\synapse_\presynaddr =  \{(\presynaddr_\ranksyn, \postsynaddr_\ranksyn, \synapticweight_\ranksyn, \synapticdelay_\ranksyn) \| \presynaddr_\ranksyn=\presynaddr\}_{\ranksyn \in [1,\Nsyn]}$. Following this definition, an event stream which evokes neurons in the presynaptic address space is multiplexed by the synapses into a new event stream which is defined by the union of the sets generated by each emitting field from the presynaptic space: 
$ \cup_{\arank \in [1,\numevent]}  \{(\postsynaddr_\ranksyn, \synapticweight_\ranksyn, \timev_\arank + \synapticdelay_\ranksyn) \}_{ \ranksyn \in \synapse_{\presynaddr_\arank}} $. In biology, this new stream of events is naturally ordered in time as events reach the soma of post-synaptic neurons. 
Synchronous activation of postsynaptic neurons, where multiple spikes converge on the soma simultaneously, will increase the firing probability of those neurons.

\begin{figure}
  \includegraphics[width=.50\linewidth]{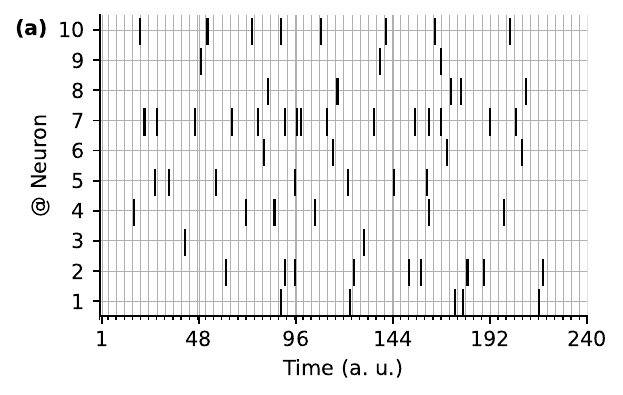}
  \includegraphics[width=.50\linewidth]{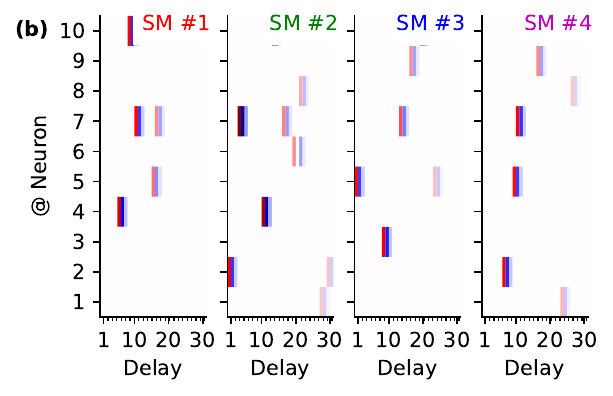}
  \\
  \includegraphics[width=.50\linewidth]{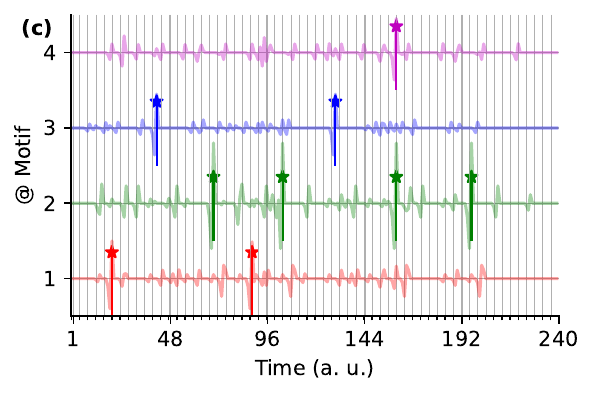}
  \includegraphics[width=.50\linewidth]{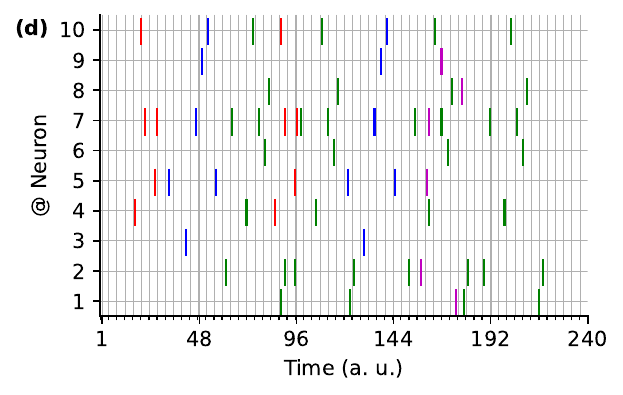} 
\caption{\textbf{From generating raster plots to inferring spiking motifs}. \textit{(a)}~As an illustration for the generative model, we draw a multiunit raster plot synthesized from $4$ different spiking motifs and for $10$ presynaptic neurons. \textit{(b)}~We show these motifs, each identified at the top by a different color. The evidence of activation (red) or deactivation (blue) is assigned to each presynaptic neuron and $31$ different possible delays. \textit{(c)}~The activation in time of the different motifs (denoted by stars) is drawn at random and then used to generate a raster plot on the multi-unit address space (see panel a). By inverting this model, an inference model can be defined for their efficient detection, outputting an evidence value (continuous line) from which the identity and timing of SMs can be inferred (vertical bars). \textit{(d)}~The original raster plot can be annotated with each identified spiking motif (as represented by the respective color assigned to SMs).
}
\label{fig:THC}
\end{figure}

From the perspective of simulating such event-based computations on standard CPU- or GPU-based computers, it is useful to transform this event-based representation into a dense representation. Indeed, we may transform any event-based input as the boolean matrix $A \in \{0, 1 \}^{N\times T}$, where $N$ is the number of presynaptic neurons in $\presynaddrspace$ and $T$ is the number of time bins (see Figure~\ref{fig:THC}a). In this simplified model, we will consider that heterogeneous delays are integers limited in range between $0$ and $D$ (that is, $\forall {\ranksyn \in [1,\Nsyn]}$, $0 \le \synapticdelay_\ranksyn < D$) such that the synaptic set can be represented by the dense matrix $\kernel^\postsynaddr \in \mathbb{R}^{N\times D}$ giving for each neuron $\postsynaddr$ the weights as a function of presynaptic address and delay (see Figure~\ref{fig:THC}b). It is equal to zero except on synapses: $\forall {\ranksyn \in \synapse^\postsynaddr}, \kernel^\postsynaddr(\presynaddr_\ranksyn,  \synapticdelay_\ranksyn) = \synapticweight_\ranksyn$. Equivalently, one may define for each presynaptic neuron $\presynaddr$ the emitting kernel as the transpose kernel $\kernel^T_\presynaddr \in \mathbb{R}^{M\times D}$, where $M$ is the number of postsynaptic neurons, whose values are zero except on synapses:  $\forall {\ranksyn \in \synapse_\presynaddr}, \kernel^T_\presynaddr(\postsynaddr_\ranksyn,  \synapticdelay_\ranksyn) = \synapticweight_\ranksyn$.
\subsection{A generative model for raster plots}

As described in Figure~\ref{fig:izhikevich}, a spiking motif can be detected using a properly tuned HD-SNN that maximizes spike synchronization at the postsynaptic terminal. Taking the argument the other way around, one may form a generative model for realistic raster plots in which spikes in the presynaptic address space are generated as the conjunction of spiking motifs defined in the postsynaptic space, knowing that both populations are connected by a set of weights and delays whose structure is stable relatively to the coding timescale. When connection weights are strong and sparsely distributed, this firing will robustly cause a specific temporal motif. Overall, these examples show that raster plots may be considered as a mixture of the effects of different elementary causes, and that each event triggers a specific spatio-temporal spiking motif. 

Formally, the activation of spiking motifs can occur independently and at random times. The activity is represented as a boolean matrix $B\in \{0, 1\}^{M\times T}$, where $M$ is the number of different spiking motifs (see Figure~\ref{fig:THC}c). Each entry $B(\postsynaddr, t)$ indicates whether a particular motif $\postsynaddr$ is activated at time $t$. The firing of a neuron $\presynaddr$ at time $t$ is considered a Bernoulli trial with a bias parameter $p(\presynaddr, t) \in [0, 1]$. This bias is conditioned by the presence of spiking motifs on postsynaptic neurons with corresponding delays. 
Assuming that this bias is conditioned by the presence of spiking motifs on \emph{all} efferent postsynaptic neurons with the corresponding delays, it can be shown that the logit (inverse of the sigmoid) of this probability bias can be written as the sum of the logit of each of these factors, whose values we will define as the corresponding weights in the kernel. We can thus write the probability bias $p(a, t)$ as the accumulated evidence given these factors as 
\begin{equation*}
p(\presynaddr, t) = \sigma\big(\kernel_\presynaddrspace(\presynaddr) + \sum_{\postsynaddr \in \synapse_\presynaddr, 0 \le \synapticdelay \le D} B(\postsynaddr, t+\synapticdelay) \cdot  \kernel_\presynaddr(\postsynaddr, \synapticdelay)  \big)  
\end{equation*}
where $\sigma$ is the sigmoid function. We will further assume that kernel's weights are balanced (their mean is zero) and that $\kernel_\presynaddrspace$ is a bias such that $\forall \presynaddr, t$, $\sigma(\kernel_\presynaddrspace(\presynaddr))$ is the average background firing rate. 

Finally, we obtain the raster plot $A\in \{0, 1\}^{N\times T}$ by drawing spikes using independent Bernoulli trials based on the computed probability biases $A \sim \mathcal{B}(p)$. Note that, depending on the definition of kernels, the generative model can model a discretized Poisson process, generate rhythmic activity or more generally propagating waves. This formulation thus defines a simple generative model for raster plots as a combination of independent spiking motifs.  This generative model can be easily extented to include a refractory period in order to ensure that there is a minimum time gap between successive action potentials, preventing them from overlapping. This temporal separation allows for discrete and well-defined neural signals, enabling accurate information processing and mitigating signal interference. The refractory period contributes to energy efficiency in neural systems and plays a crucial role in temporal coding by creating distinct time windows between successive spikes. 
\subsection{Detecting spiking motifs}
%
Assuming the spiking motifs (as defined by the kernel $\kernel$) are known, the generative model allows to determine an inference model for detecting sources $\hat{B}$ when observing a raster plot $A$. Indeed, by using this forward model, it is possible to estimate the likelihood $p(b, t)$ for the presence of a spiking motif of address $b$ and at time $t$ by using the transpose convolution operator. This consists in using the emitting field $\synapse_\presynaddr$ of presynaptic neurons in place of the receptive field $\synapse^\postsynaddr$ of postsynaptic neurons. It thus comes that when observing $A$, then one may infer the logit of the probability as the sum of evidences:
\begin{equation*}
  p(\postsynaddr, t) = \sigma\big(\kernel_\postsynaddrspace(b) + \sum_{\presynaddr \in \synapse^\postsynaddr,  0 \le \synapticdelay \le D} A(\presynaddr, t-\synapticdelay) \cdot \kernel^\postsynaddr(\presynaddr, \synapticdelay) \big)  
\end{equation*}
This also takes the form of a temporal convolution. This assumption holds as long as the kernels are uncorrelated, a condition which is met here numerically by choosing a relatively sparse set of synapses (approximately $1\%$ of active synapses). Finally, we compute $\hat{B}$ by selecting the most likely items, allowing to identify the spiking motifs in the input raster plot (see Figure~\ref{fig:THC}d). 

One may naturally extend this algorithm when the spiking motifs (that is, the weights) are not known, but that we know the timing and identity of the spiking motifs. Indeed, the equation above is differentiable. Indeed, the activation function of our spiking neural is a sigmoid function implementing a form of  Multinomial Logistic Regression (MLR)~\cite{grimaldi_learning_2023}.  
The underlying metric is the binary cross-entropy, as used in the logistic regression model. In particular, if we consider kernels with similar decreasing exponential time profile, one can prove that this detection model is similar to the method of Berens {\it et al.}~\cite{berens_fast_2012}. In our specific case, the difference is that the regression is performed in both dendritic and delay space by extending the summation using a temporal convolution operator. 
\section{Results}
To quantify the efficiency of this operation, we generated raster plots parameterized by $N=128$ presynaptic inputs and $M=144$ synthetic spiking motifs as random independent kernels and with $D=31$ possible delays. We drew random independent instances of $B$ with a length of $T=1000$ time steps and an average of $1.0$ spikes per neuron. This allowed us to generate a large number of synthetic raster plots, which we use to infer $\hat{B}$. We compute accuracy as the rate of true positive detections (both for inferring the address and its exact timing) and observe on average $\approx 98.8\%$ correct detections.

\begin{figure}
  \centering
  \includegraphics[width=0.335\linewidth]{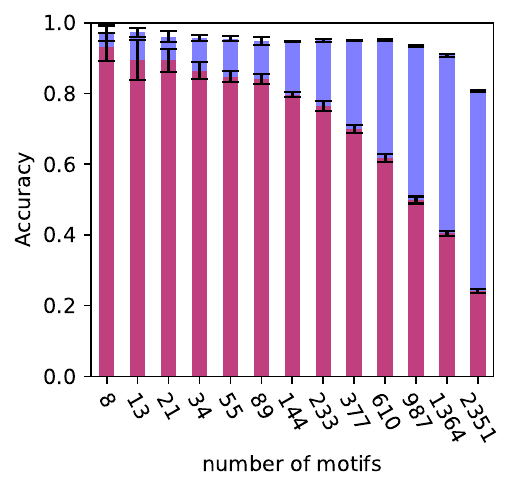}
  \includegraphics[width=0.320\linewidth]{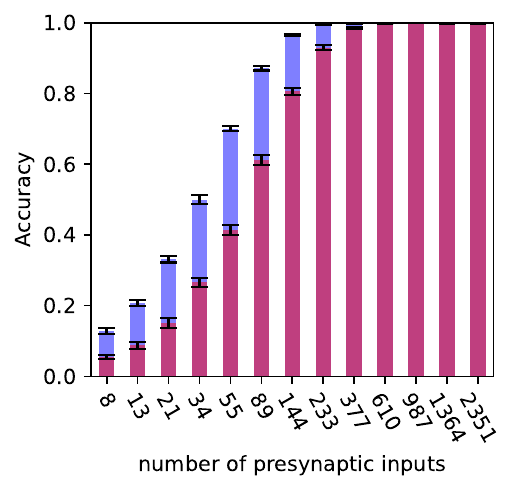}
  \includegraphics[width=0.315\linewidth]{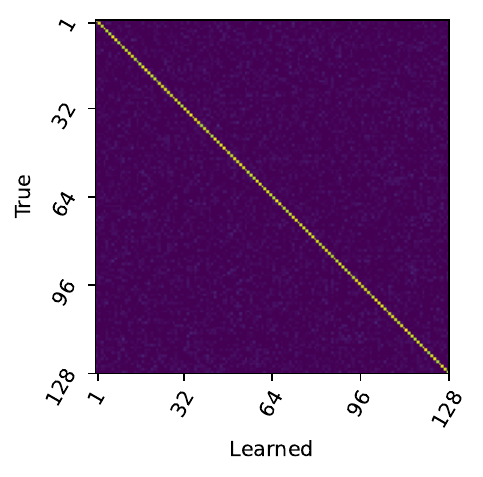}
    \caption{{\bf Detecting spiking motifs using spiking neurons with heterogeneous delays.} 
Accuracy of detection for the classical correlation (red) and the HD-SNN method (blue) as a function of  {\it (Left)}~the number $M$ of kernels, 
    {\it (Middle)}~the number of presynaptic neurons, 
    {\it (Right)} Correlation matrix of true vs learned kernels.
    }
  \label{fig:model_results}
\end{figure}

We extended this result by showing how accuracy evolves as a function of the number of simultaneous spiking motifs, holding the frequency of occurrence constant. We show in \fig{model_results}~(left) that the accuracy of finding the right spiking motif is still above $80\%$ accuracy with more than $1364$ overlapping spiking motifs. This observation illustrates quantitatively the capacity  of the HD-SNN in representing a high number of simultaneous motifs. Furthermore, we show in \fig{model_results}~(middle) that (with $M=144$ spiking motifs fixed) the accuracy increases significantly with increasing temporal depth $D$ of the spiking motif kernel, quantitatively demonstrating the computational advantage of using heterogeneous delays. These results were obtained under the assumption that we know the spiking motifs through $\kernel$. However, this is generally not the case, for example, when considering the raster plot of biological neurons.

Finally, we evaluated the performance of the supervised learning scheme in inferring the connection kernel when the address and timing of spiking motifs are known. The kernel was initialized with random independent values, and we used stochastic gradient descent with a learning rate of \num{1e-4} over $\num{1e4}$ trials (i.e., over rasters as defined above with $T=1000$ and $N=128$). Qualitatively, the convergence was monotonous, and the correct values of the $M=144$ spiking motifs were quickly recovered. Quantitatively, the correlation between the true and learned kernel weights showed that all kernels were correctly recovered (see Figure~\ref{fig:model_results}, right). Performing inference with the learned weights was as efficient as with the true kernels, and showed no significant difference (not shown).

%
\section{Discussion}
\subsection{Synthesis and Main Contributions}
In this paper, we present a novel Heterogeneous Delays Spiking Neural Network (HD-SNN) model designed for the detection of spiking motifs in synthetic neurobiologically-inspired raster plots.

Our contributions encompass several innovations. Firstly, we formulate the HD-SNN model from first principles, optimizing the detection of event-based spatiotemporal motifs. Unlike previous models like the tempotron, which are evaluated on simplified problems, our model is rigorously tested on realistic synthetic data. The results demonstrate that, assuming that the spiking motifs are known, our model accurately detects the identity and timing of spiking motifs, even when multiple motifs are superimposed. Additionally, we show that our method outperforms correlation-based heuristics, such as those used in previous works like~\cite{ghosh_spatiotemporal_2019,yu_stsc-snn_2022}, in terms of efficiency. Secondly, compared to other event-based methods, like HOTS~\cite{lagorce_hots_2017}, our model's weights are interpretable. These weights are directly related to the logit, which is the inverse sigmoid of the probability of detecting each spatiotemporal spiking motif. Finally, a crucial novelty lies in the simultaneous learning of weights and delays in our model. In contrast, models like the polychronization model~\cite{izhikevich_polychronization_2006} only learn weights and delays are frozen. These contributions highlight the significance and effectiveness of our HD-SNN model for detecting spiking motifs, offering insights into the neural mechanisms involved in pattern recognition and information processing.

\subsection{Main limits}

The model comes with certain limitations. First, the entire framework is based on discrete time binning, which is incompatible with the continuous nature of biological time. While this choice facilitated efficient implementation on conventional hardware such as GPUs, it can be extended to a purely event-based SNN framework~\cite{grimaldi_robust_2023}. By analytically incorporating a precision term in the temporal value of the input spikes, a purely event-based scheme can be achieved, promising speedups and computational energy gains.

Second, the current model is purely feed-forward, i.e. the spikes generated by postsynaptic neurons are based solely on information from their classical receptive fields. However, neural systems often involve lateral interactions between neurons in the same layer and feedback connections, which can be crucial for computational principles and modulation of neural information. While our theoretical model can incorporate these recurrent connections by inserting new spikes into the list of spikes reaching presynaptic addresses, it requires proper tuning to avoid perturbations of the homeostatic state. For the implementation of predictive or anticipatory processes, recurrent activity would be essential, especially when dealing with multiple different delays that require temporal alignment. Such recurrent activity has previously been modelled to explain phenomena such as the flash-lag illusion. Implementing this using generalised coordinate and delay operators would allow predictive mechanisms to be incorporated into our proposed HD-SNN model, providing an elegant solution to this problem.

Addressing these limitations and exploring the extension of the HD-SNN model to event-based schemes and recurrent connections would enrich its potential applications and pave the way for a better understanding of neural information processing in complex systems.
\subsection{Perspectives}
The coding results were obtained under the assumption that we know the spiking motifs by way of $\kernel$, or using supervised learning by knowing the identity and timing of spiking motifs. However, this is generally not the case, e.g. when observing the neurobiological raster plot of a population of neurons. One perspective would be to extend the model to a fully self-supervised learning paradigm, i.e. without any labeled data~\cite{barlow_unsupervised_1989}. This type of learning is thought to be prevalent in the central nervous system and, assuming the signal is sparse~\cite{olshausen_emergence_1996}, one could extend these Hebbian sparse learning schemes to spikes~\cite{perrinet_emergence_2004,masquelier_competitive_2009}. 

We expect that this would be particularly adapted for exploring neurobiological data~\cite{mackevicius_unsupervised_2019}. Indeed, there is a large literature showing that brain dynamics often organize into stereotyped sequences such as synfire chains~\cite{ikegaya_synfire_2004}, packets~\cite{luczak_sequential_2007}, or hippocampal sequences~\cite{villette_internally_2015} (for a review, see~\cite{grimaldi_precise_2023}). These motifs are stereotyped and robust, as they can be activated in the same motif from day to day~\cite{haimerl_internal_2019}. In contrast to conventional methods used to process neurobiological data, such an event-based model would be able to answer key questions regarding the representation of information in neurobiological data. 
%

\end{document}